\documentclass[aps,prl,preprint,superscriptaddress]{revtex4-1}
\usepackage[final]{graphicx}
\usepackage{epstopdf}
\usepackage{subfigure}
\begin{document}

%\preprint{}

\title{Liquidity crises on different time scales}


\author{Francesco Corradi}
\affiliation{``Sapienza'' University of Rome, Italy}
\author{Andrea Zaccaria}
\affiliation{ISC-CNR, Italy}
\affiliation{``Sapienza'' University of Rome, Italy}
\author{Luciano Pietronero}
\affiliation{``Sapienza'' University of Rome, Italy}
\affiliation{ISC-CNR, Italy}
%\thanks{}


\date{\today}

\begin{abstract}
We present an empirical analysis of the microstructure of financial markets and, in particular, of the static and dynamic properties of liquidity. We find that on relatively large time scales (15 minutes) large price fluctuations are connected to the failure of the subtle mechanism of compensation between the flows of market and limit orders: in other words, the missed revelation of the latent order book breaks the dynamical equilibrium between the flows, triggering the large price jumps. On smaller time scales (30 seconds), instead, the static depletion of the limit order book is an indicator of an intrinsic fragility of the system, which is related to a strongly non linear enhancement of the response. In order to quantify this phenomenon we introduce a measure of the liquidity imbalance present in the book and we show that it is correlated to both the sign and the magnitude of the next price movement. These findings provide a quantitative definition of the effective liquidity, which results to be 
strongly dependent on the considered time scales.
\end{abstract}

%\keywords{}

\maketitle
\section{Introduction}
The application of tools and methods borrowed from the fields of statistical mechanics and the physics of complex systems to the study of financial markets has led to a number of empirical and theoretical results \cite{bouchaud,mantegnabook}. In particular, the analysis of markets' microstructure \cite{harris,has} benefits from the huge amount of available data, which permits to ground models and findings on a strong empirical basis \cite{farbouch}. One of the most important issues is related to the subtle relationships among order flow, liquidity and price movements. The empirical evidence, in fact, is in contrast with the natural guess that price movements should be trivially related to the arrival of orders and, in particular, to their volume. Firstly, orders are strongly correlated in sign \cite{lobato,fhfhfhfhfh,subbtle}, in sharp contrast with the well known absence of autocorrelation in price returns, which is at the basis of the efficient markets hypothesis \cite{fama}. Moreover, the magnitude of 
price movements has been shown to be little dependent on the volume of the incoming orders \cite{weber2,weber1,volvol}. The possibility to reconstruct the microstructure of financial markets, namely, the so called limit order book (LOB) \cite{biais,gould2013limit}, permits to stress the importance of agents' strategies on the state of the market, for example, on spread dynamics and volatility clustering \cite{zaccaria2010asymmetric}, and the reaction in terms of balancing order flow to explain the magnitude of price movements \cite{Chordia2002111,farmer6,stephens,ofi}. These findings directly lead us to the concept of \textit{liquidity}. Even if it is widely recognized as ``the most important characteristic of well-functioning markets"\cite{harris}, liquidity does not have a single shared definition. Generally speaking, a market is \textit{liquid} if agents can trade large volumes quickly, with little impact on the price and facing low transaction costs. In other words, it is a measure of the ease of converting assets into 
legal tender (which is, by definition, 
fully liquid) and viceversa. From a practical point of view, and keeping in mind possible applications to the LOB, this situation can be described by a number of features which make a market liquid. Using the suggestions of Sarr and Lybek \cite{sarr2002measuring}, we can say that a market is liquid if it has the following characteristics: 
i)Tightness: both implicit and explicit transaction costs are low. ii) Immediacy: the trading speed is high. iii) Depth: the orders already placed in the book cover a wide range of prices. iv) Breadth: the best quoted orders are large enough to guarantee a low impact even for large incoming orders. v) Resilience: the market reacts quickly to possible order imbalances by means of an opposite order flow \footnote{These features could be referred to the whole market, but in the following we will focus not on market liquidity in general but on the liquidity of a single stock.}. These features are linked, even if they capture different aspects of liquidity. In this work we will argue that the relative importance of these different aspects depends on the time scales in play and, in particular, we will focus on the \textit{depth} and the \textit{breadth} of the LOB for small time scales and on its \textit{resilience} for large time scales, introducing suitable measures of liquidity. The importance of the latter is 
based on a number of empirical studies on the LOB, which singled out liquidity as the main driver of large price fluctuations \cite{farmer8}. In particular, the distribution of returns due to single transactions has been proved to be substantially equal to the distribution of the first gaps of the LOB\cite{farmer6}. Toth et al. \cite{toth2009studies} carried out an extensive study on the post-large events dynamics which characterizes large price changes, finding evidence for power law relaxations typical of complex systems \cite{ponzi2009market}.\\
In our work we will study which kind of LOB dynamics may eventually lead to large price changes. The aim is to identify those situations in which an intrinsic state of instability enhances the probability to have a price jump on the short run. The first step is to build a methodology to identify large events in an unbiased way. Then, we study the order flows around these events, finding peculiar patterns, especially at the best prices. The analysis of the statics and the dynamics of the LOB permits to define in a quantitative way liquidity, whose imbalance is directly related to future price changes. Our focus will change with the time scales in play. In particular, order flow will turn to be the main driver on large time scales while the depletion of one side of the book will turn to be more important on small time scales. In general, this empirical work confirms the idea that price changes, and in particular large fluctuations, are closely related to liquidity dynamics. To give the reader an idea of the 
potential role of liquidity we cite the fact that, for a large cap stock, the typically available liquidity is about $1\%$ of the daily traded volume, which is, in turn, only $0.1\%$ of the total capitalization \cite{bouchaud2010endogenous}.\\
\begin{figure}[!ht]
\centering
\includegraphics[scale=0.45]{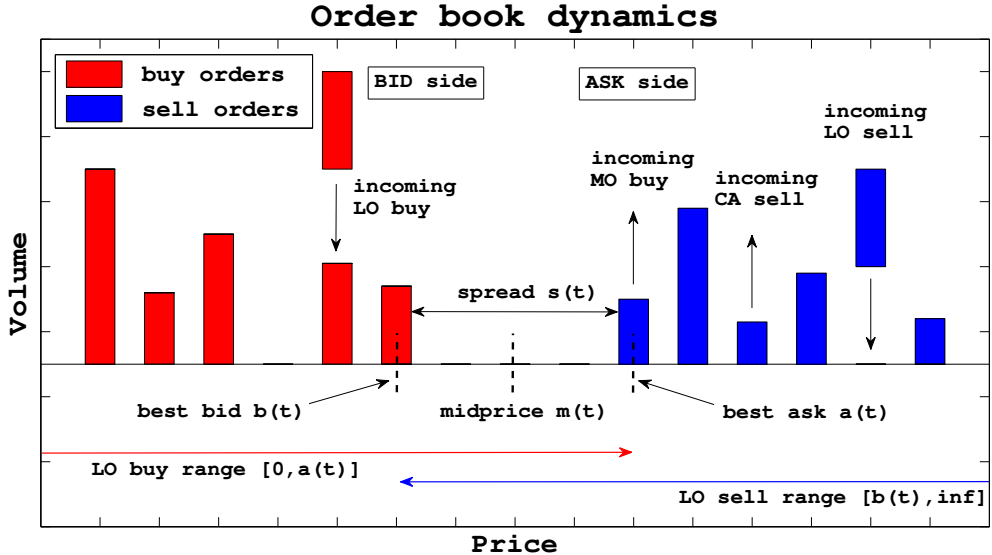}
\caption{Synthetic view of the different operations available on the electronic market. Incoming limit orders add liquidity to the book, while market orders and cancellations remove such availability of orders. The midprice $m(t)$ can change if, as a consequence of these operations, the best sell limit orders (ask) or the best buy limit orders (bid) change.}
\label{LOBworks}
\end{figure}
Before proceeding let us briefly recall how a LOB works. The different elements we are going to introduce are depicted in Fig.\ref{LOBworks}. Traders have at their disposal three different \textit{operations}: i) they can place a \textit{limit order} (LO), which is the proposal to buy or sell a certain amount (or \textit{volume}) of shares at a fixed price. Since limit orders are usually not immediately traded, they are stored in the book until they can be matched with another order. ii) they can decide to buy or sell a volume of shares at the best available price (these prices are called best \textit{ask} and best \textit{bid}, respectively), that is, place a so called \textit{market order} (MO) which will be matched with the best available limit order. iii) \textit{cancel} (C) a placed LO. Potentially all these three operations may have an impact on the price, which is usually defined as the value at which the last transaction occurred or the average between best bid and best ask (in this last case the 
price is called \textit{midprice}). For example, an incoming MO to buy a given volume of shares could remove totally the best ask, triggering a price increment. Traders can place sell (buy) LOs also at a price which is lower (higher) than the best bid (ask): in this case one 
can see all or part of the LO as an \textit{effective} MO, because it will trigger an immediate transaction. In the following, as it is usually done in the literature, we will call MOs the effective MOs. The price of a LO has to be chosen among a discrete set of prices. For example, a LO could be placed at 9.5 or 9.75 dollars, but not at 9.66. The minimum distance in price between two orders is called \textit{tick}. In the previous example, the tick was equal to 0.25. This is the natural measure of distance among the placed limit orders, which constitute the limit order book.\\
Our empirical analysis is based on a database of various stocks traded at the London Stock Exchange. In particular, we analyzed four liquid stocks (AZN, BP, RBS, and VOD) but we show, for reasons of space, only the results for AZN. We point out that we performed our analysis also on the other three stocks, finding similar results. Our database covers the whole year 2002. As it is usually done in the literature, we discarded the first 30 minutes of each trading day to avoid anomalous effects due to the opening of traders' positions.
\section{Unbalanced order flows during large price fluctuations}
\subsection{Selection of large events}
In this section we will be mainly interested in large price fluctuations, so we need in the first place a criterion to define a \textit{large event}. A naive approach would select those time windows within which a return above a certain fixed threshold, let us say $x\%$, occurred. However, such an \textit{absolute} filter would be affected by the intraday volatility pattern and will select mostly events from the begin and from the end of the trading day. On the other hand a \textit{relative} filter, in which one selects those events which are larger than the average fluctuations at the same time of the day, will suffer from the effect of an opposite uneven selection. As a consequence, in this work we use a combination of the two filters, following \cite{zawadowski2004large,RePEc:taf:quantf:v:6:y:2006:i:4:p:283-295,toth2009studies}. If one chooses suitable thresholds for both filters an uniform distribution of events over the day can be recovered.\\
The authors of the above mentioned papers do not set a time interval for the large event, because a large return which passes both filters may be defined on different time scales. So they simply set the temporal reference frame in order to have $t=0$ when both filters are passed. In this way, they can study only the post-event dynamics, since events have different temporal dimensions. Moreover, also a classification as a function of the return is troublesome: since the algorithm which searches for the large events stops when the thresholds are reached, all events will tend to have similar returns, since the only possible variation will be due to the U-shaped intraday volatility pattern. In order to avoid these drawbacks we use time windows of fixed length $\Delta t=15$ min. In practice, for every operation in the LOB we check if a large price fluctuation is present in the next $\Delta t$. The specific value of $\Delta t$ is chosen to have a reasonable compromise between two needs: to have enough statistics 
the time interval can not be too large, but to be able to study the dynamics inside the event $\Delta t$ can not be too small. Obviously, by keeping the time windows' size constant we are treating in the same way events that could be very different, the only thing in common being the presence of a large net return between the start and the end of the considered time interval. However this is the only way to make a pre-event analysis possible and to have a broad distribution of returns to analyze. We would like to stress that the heterogeneity of events is an intrinsic feature of this kind of analysis: for example, using the same criteria of \cite{toth2009studies}, one could find events with a temporal extension up to 2 hours. An important issue is the presence of strong autocorrelations in almost every time series of interest and, in particular, the volumes of incoming orders and the absolute values of returns. To avoid our analysis to be undermined by such effects we discard all those events with another 
close event before, that is, we discard those events which are preceded by another large event at a distance $\Delta t$.\\
In practice, for the (relatively) large time scales analysis ($\Delta t=15$ min) we select those time windows with a price fluctuation larger than $x=0.5\%$ (in absolute value) and larger than $n=3$ times the average volatility $\sigma$ during that time of the day. After the application of the various filters we find 332 positive and 362 negative events for AZN, 358 positive and 381 negative events for BP, 306 positive events and 325 negative events for RBS, 402 positive and 436 negative events for VOD.
\subsection{Order flows before and during large fluctuations}
\begin{figure}[!ht]
\centering
\subfigure[]
{\includegraphics[scale=0.55]{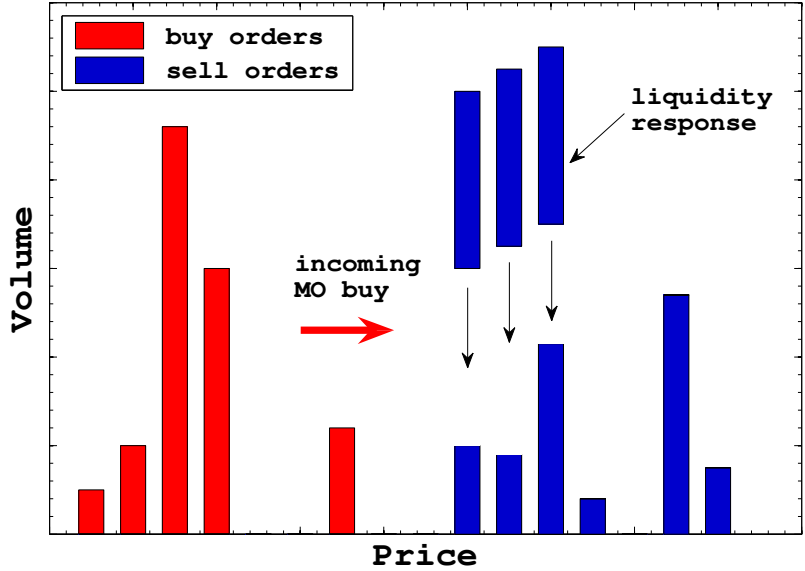}}
\hspace{5mm}
\subfigure[]
{\includegraphics[scale=0.6]{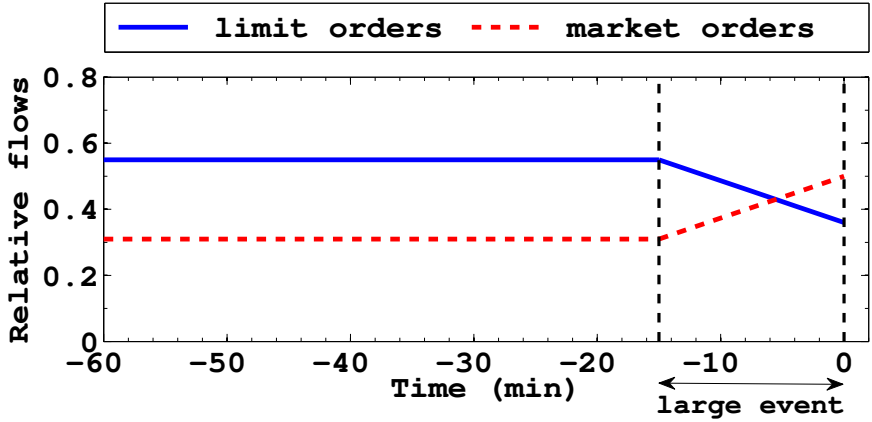}}
\caption{(a): Sketched representation of the reaction, in terms of placement of limit orders, to an incoming flow of market orders. (b): We have a large event when the relative flows change, breaking the dynamical equilibrium.}
\label{resilience}
\end{figure}

\begin{figure}[!ht]
\centering
\includegraphics[scale=0.4]{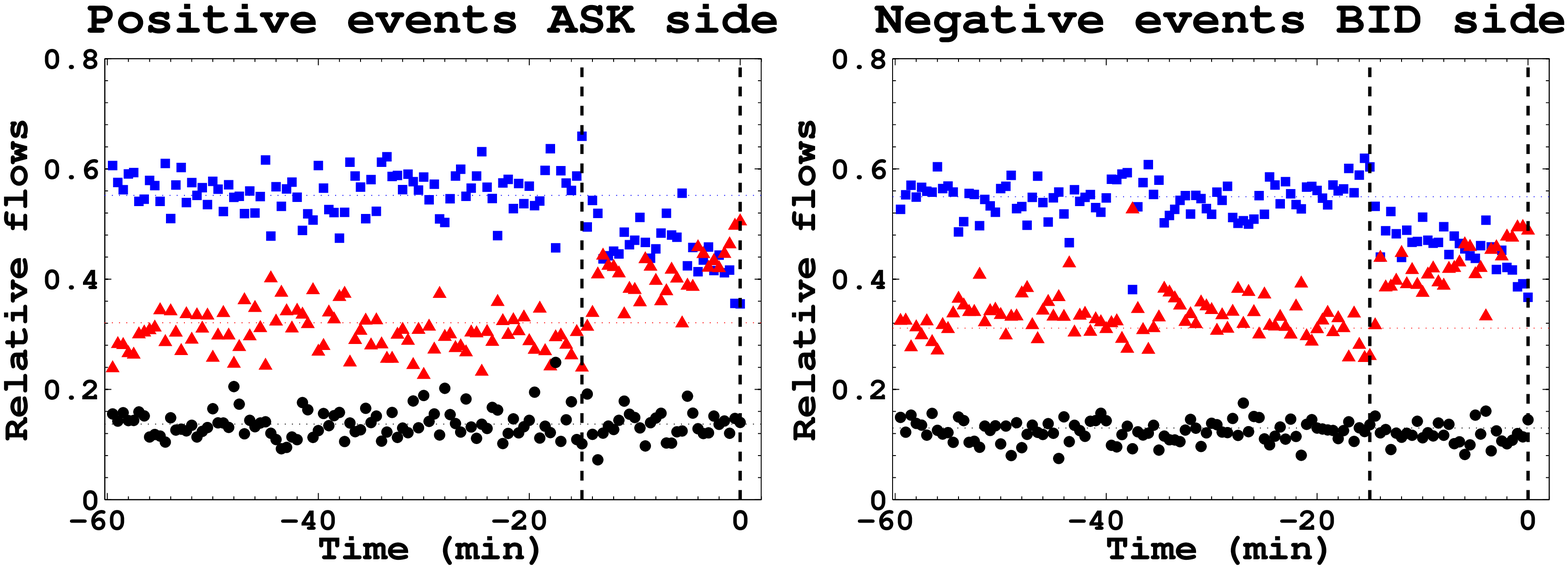}
\includegraphics[scale=0.4]{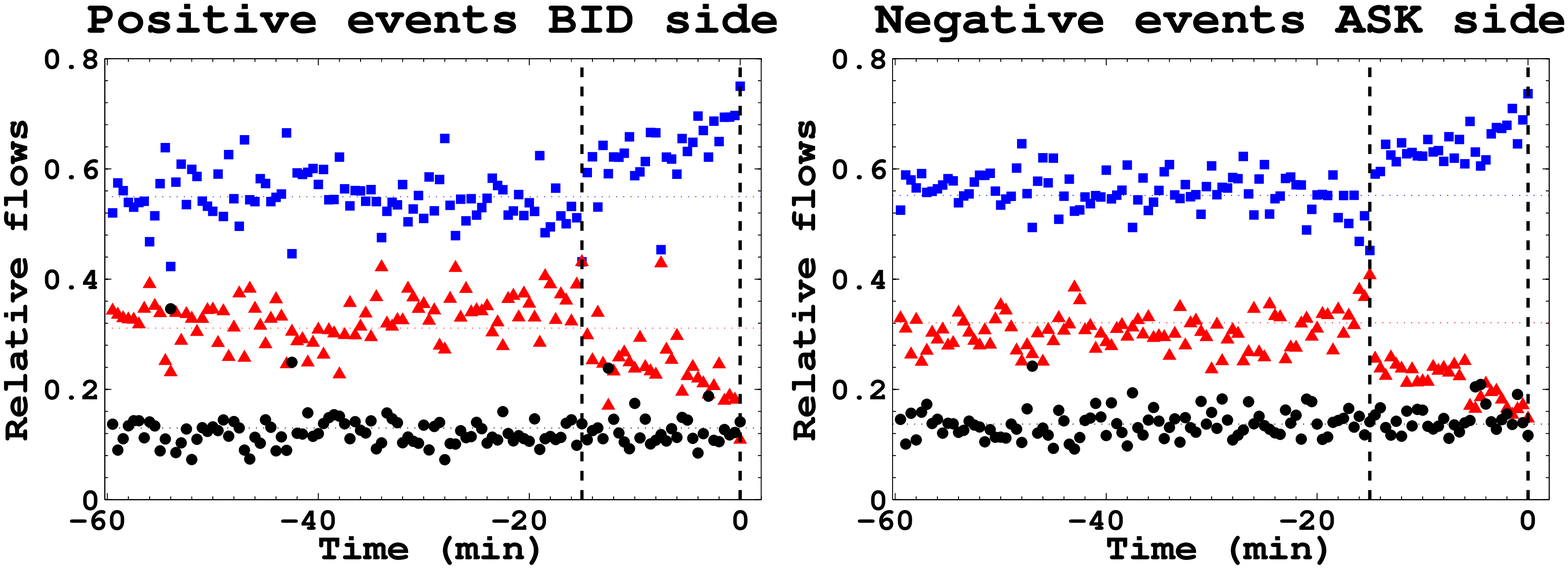}
\caption{Relative volumes of operations at the best price before and during large price fluctuations, defined in time windows of 15 minutes. On top, the two sides under pressure. The increase of market orders is not balanced by an opposite response of limit orders. At the bottom, the pressing side experiences the opposite situation. Cancellations remain constant in all four cases.}
\label{Flows_rates}
\end{figure}

It is a well known feature of LOBs that the arrival of a significant flow of MOs triggers a liquidity response in terms of LOs, which are usually placed close to the best price of the pushed side \cite{stephens,toth2012does}. For example, if a persistent flow of MOs, such as the ones that result from the splitting of a large hidden order \cite{PhysRevE.80.066102}, is somehow detected, the other traders will likely provide more liquidity, effectively contrasting the incoming flow (this behavior is sketched in Fig.\ref{resilience}(a)). As a consequence, the ratio of MOs to LOs remains constant even if the two volumes can fluctuate a lot, keeping the market resilient and contrasting large price fluctuations. In this section we will see that the break of this dynamical equilibrium is a key element of large price fluctuations. In fact, we will show that during large price movements the flow of LOs is no more able to stand in the way of the incoming MOs, as schematically depicted in Fig \ref{resilience}(b). We 
note that, in our 
reasonings, cancellations have not been mentioned, even if their volume is not negligible. This is due to the fact that in the analyses we are going to present their relative flow does not change from normal times to time windows characterized by large price movements.\\
In order to check this line of reasoning, we empirically study the orders' flow before and during the large events we defined in the previous section. First of all we point out that during large events all flows exhibit large fluctuations. In particular, the side under pressure experiences an increase of LO and cancellations up to about three times, and of MO up to five times, with respect to the respective annual averages. In the following we will consider the relative flows of the three possible operations (LO, MO and cancellations) with respect to their sum during large events. Since we will study one side of the book at a time and positive and negative large events separately, we will have four different relative flows for each operation. For example, the relative flow of LOs on the ask side, denoted by $A$, during large positive events, denoted by $+$, is calculated in the following way:
\begin{equation}
 r^{A+}_{LO}=<\frac{Q_{LOS}}{Q_{LOS}+Q_{MOB}+Q_{CS}}>_{A+}
\label{relative}
 \end{equation}
 
where $Q_{LOS}$ is the volume of sell LOs, $Q_{MOB}$ the volume of buy MOs, and $Q_{CS}$ is the volume of the cancellations of sell LOs. So we consider the relative flows of all the operations which are made on the same side of the book during large events. The averages are performed by considering a total time range of one hour from the end of the large event, but backwards, in such a way that 15 minutes belong to the large event and 45 minutes belong to the pre-event. We divide such a time range in subintervals of 30 seconds each and we calculate the averages $<...>$ for each subinterval separately. The other relative flows are calculated in a similar way, always considering operation made on the same side of the book.
In Fig.\ref{Flows_rates} we plot the relative flows of MOs (red triangles), LOs (blue squares) and cancellations (black dots) at the best bid and ask for both positive and negative events. The beginning of the large event is indicated by the vertical dashed line, while the horizontal line represents the relative flows calculated by averaging over the whole year. While cancellations remain roughly constant in all four plots, one can notice that the side of the book which is under pressure (that is, the ask side during positive events and the bid side during negative events) shows an excess flow of MOs which is not balanced by a corresponding increase of LO flow. 
The opposite behavior is evident for the pressing side of the book. This situation is clearly related to the presence of large price fluctuations: since we have a large flow of MOs not opposed by an adequate flow of LOs, the MOs face a low resistance and can penetrate into the side under pressure of the order book, leading to a large price movement. \\ This situation is, at a first glance, in sharp contrast with the results presented in \cite{toth2009studies}, where the relative flows remain constant before and after the large event. Moreover, the relative number of MOs and cancellations are practically inverted: while in \cite{toth2009studies} one can see a relative number of cancellations roughly equal to 0.35 and a relative flow of MOs around 0.05, in Fig.\ref{Flows_rates} we observe approximately 0.32 for MOs and 0.15 for cancellations. While the discrepancy between the average values is due to our choice to show only the flows at the best bid and ask, we believe that the sharp increase of MOs and the 
decline of LOs are visible only if one 
averages over a coherent set of events. In fact, while we keep our time windows fixed, Toth et al. consider events of very different duration, and time windows which could span, in principle, from a few minutes up to two hours. Moreover, we point out that the increase of MOs and the decrease of LOs is much less evident, even if still clearly visible, when one selects the large events by means of our filter but considers the whole book instead of the bests. 
\subsection{Lack of display of latent liquidity}
\begin{figure}[!ht]
\centering
\includegraphics[scale=0.55]{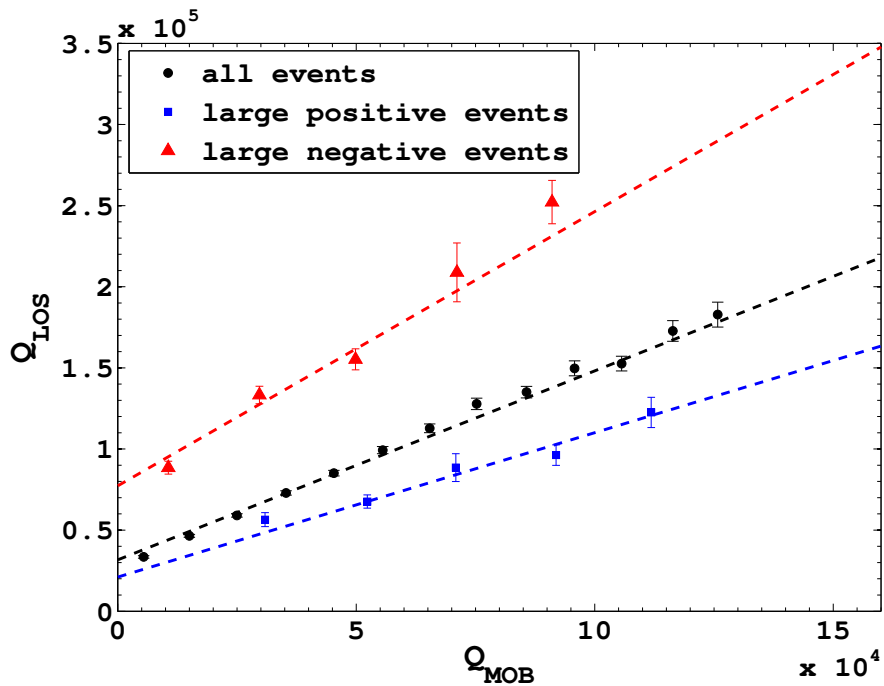}
\caption{Comparison between market and limit order flows, on the ask side of the book, on normal times and during large price movements. With respect to the average (black dots), during large positive events (blue squares) there is a lack of display of hidden liquidity, namely, too few limit orders to counterbalance the market orders flow. The opposite behavior can be seen during large negative events (red triangles). A similar analysis can be performed on the bid side of the book, finding analogous results.}
\label{response}
\end{figure}

The previous analysis suggests that large price fluctuations are due to an unexpected lack of response in terms of liquidity to an incoming MO flow, that is, a loss of \textit{resilience}. In order to confirm this view we studied how the number of LOs placed at the best price varies as a function of the number of the incoming MOs. Let us focus on the ask side of the book (the same results are valid for the bid side). Since we find a clear correlation between the volumes of MOs to buy $Q_{MOB}$ and LOs to sell $Q_{LOS}$ placed at the best ask in the same time window of size $\Delta t=15$ min, we can use, as a first order approximation, a linear relationship of the type $Q_{LOS}\approx a Q_{MOB} + b$, where $a$ and $b$ can be seen as proxies for market's resilience. We point out that at 
this stage of our analysis, being both flows calculated on the same time window, there is no clear causal relationship between the two, even if one usually thinks of the placement of LOs close to the best as a reaction to an incoming MOs flow. To make cause-effect relationships come to light we need a finer temporal resolution, which will be adopted in the following sections.\\ Now we study how the linear relationship between the volumes of MOs and LOs changes during the large price fluctuations. To do so we consider the order flows $Q_{LOS}$ and $Q_{MOB}$ in each time window, but selecting the latter in three different ways: during normal times, large positive and large negative events. We come up with three sets of points in the $Q_{LOS}$-$Q_{MOB}$ plane. In Fig.\ref{response} we plot both the parametric and the non parametric (binning) fits of the three sets of points, finding that the response in terms of sell LOs on the ask side of the book is much less than the average during large positive events. The 
black dots represent the average response, which is measured considering all time windows. The linear relationship is clear. The blue squares represent the same calculation, but considering only that subsample of windows which have been classified as large positive events, that is, the ones in which the ask side is under pressure and does not resist to the incoming flow of MOs. One can see that both the coefficient $a$ of the linear approximation and the absolute mean values of LOs, given a similar flow of MOs, are lower. On the contrary, during the large negative events (red triangles), in which the considered side of the book is pressing, one can see an excess placement of LOs. We find very similar results for the bid side of the book and the relationship between $Q_{LOB}$ and $Q_{MOS}$.\\ 
This is a direct, empirical evidence of the relationship between large price fluctuations and traders' operations. It is important to recall herein that most of the liquidity of the book is \textit{hidden}, in the sense that most traders prefer not to place their limit orders until a clear flow of market orders is detected. Only in this case, and proportionally, additional liquidity is provided. We have seen that during large events this mechanism of automatic compensation stops: a big price jump is therefore related to a lack of revealed liquidity (which is instead present during normal times) and which makes the market less resilient.
\section{Depletion of the limit order book on short time scales}
\begin{figure}[!ht]
\centering
\includegraphics[scale=0.27]{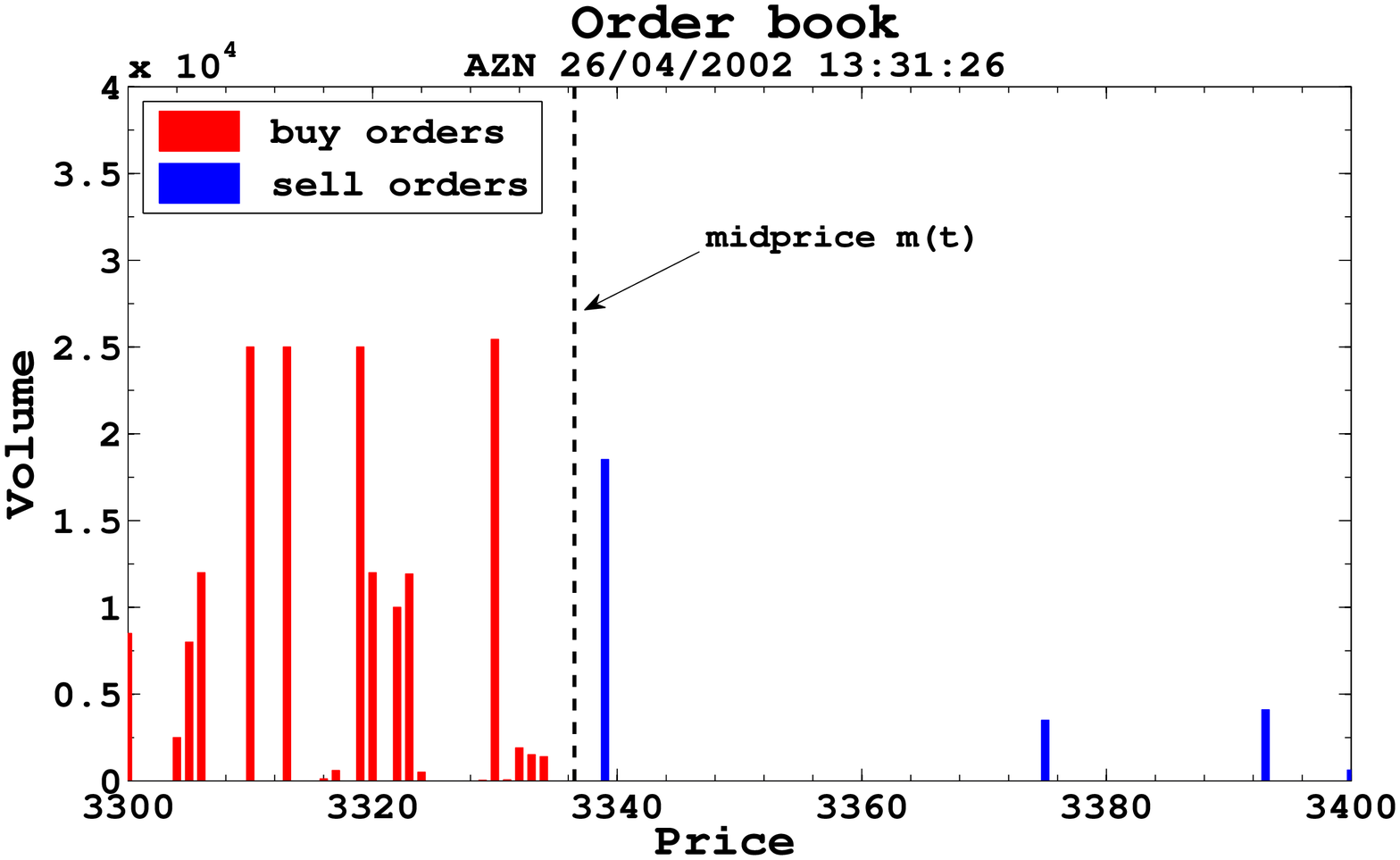}
\includegraphics[scale=0.27]{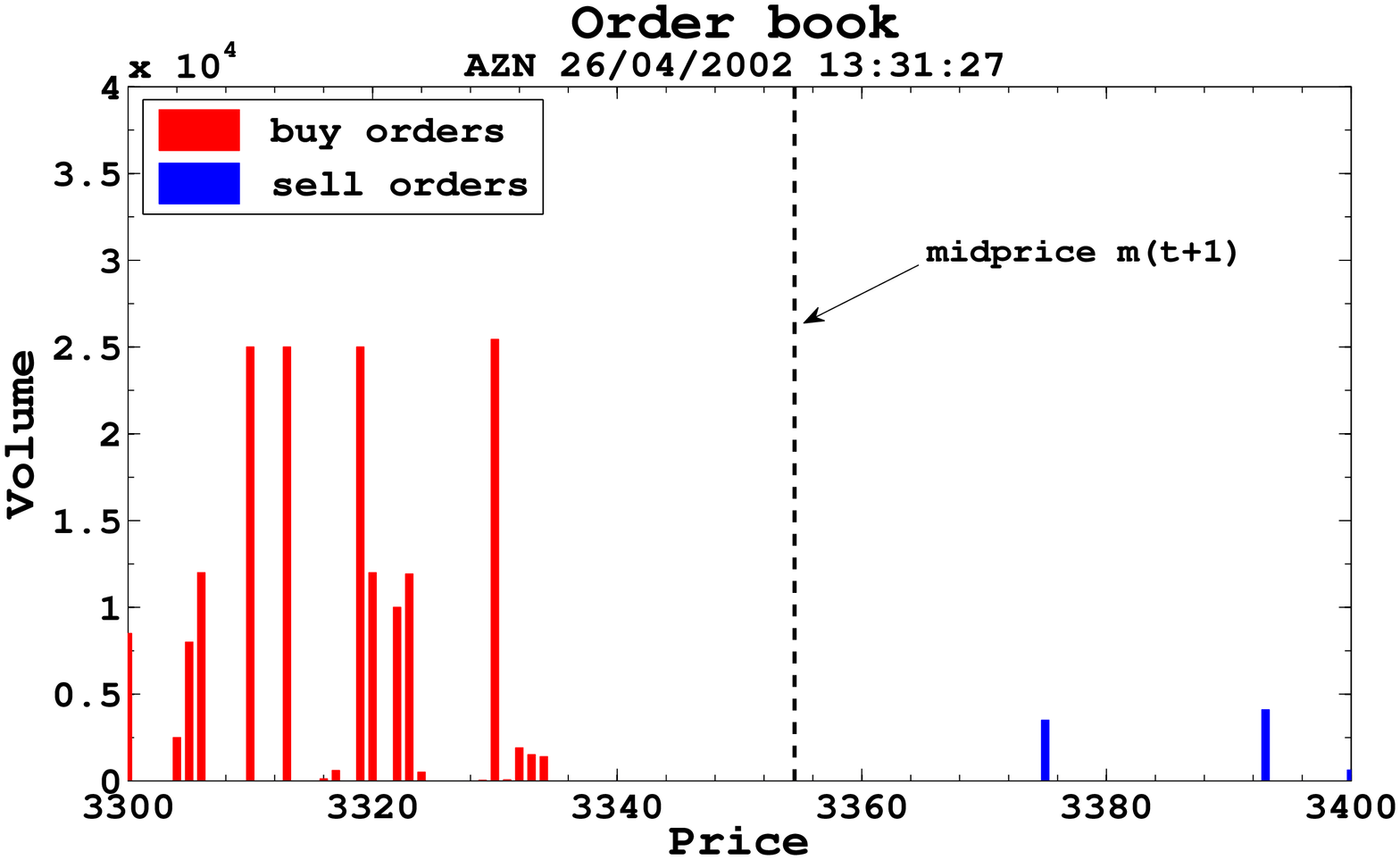}
\caption{Large price fluctuations are driven mostly by the presence of large gaps in the LOB. On the left side, a situation of intrinsic fragility of the ask side book. On the right side, the consequence of an incoming MO, which triggers a large midprice change.}
\label{sfondexample}
\end{figure}

\begin{figure}[!ht]
\centering
\includegraphics[scale=0.5]{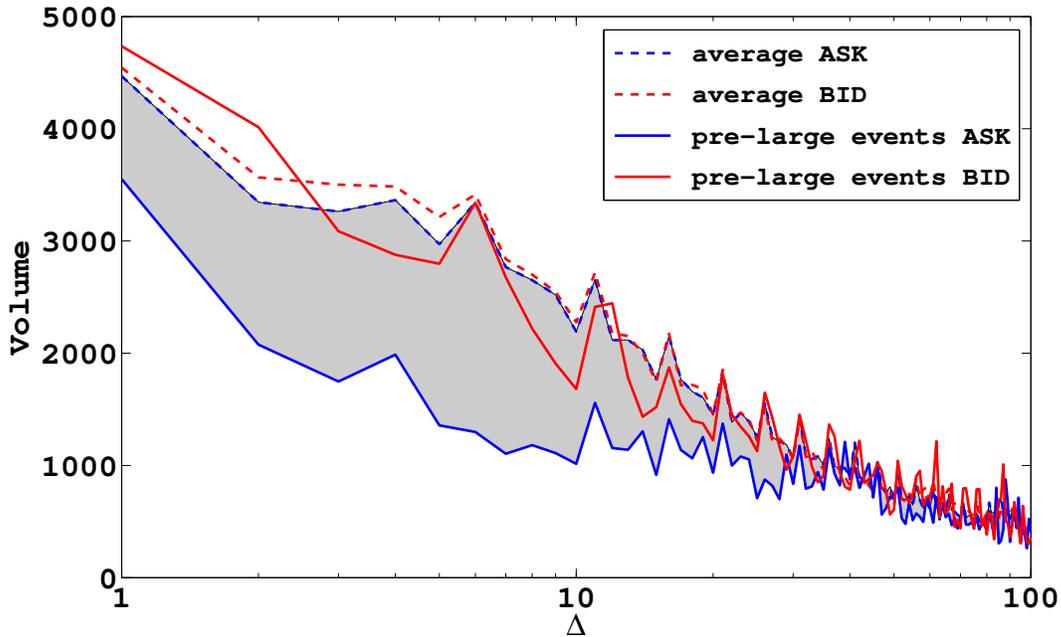}
\caption{Comparison between the average order book profile with the average profile before a large positive price fluctuation. While the bid profile before the event is roughly similar to the average ask and bid profiles, the ask profile, that is, the side which is going to break down, is much less liquid.}
\label{OBdepl}
\end{figure}

The main empirical properties of financial markets, the so called stylized facts \cite{bouchaud,mantegnabook}, are present on very different time scales. In particular, the logreturns $r_{\Delta t}=\log p(t+\Delta t) - \log p(t)$ are power law distributed for values of $\Delta t$ ranging from seconds to months. Only if one considers single operations this scale invariance breaks down, as expected, because of discreteness effects. In this section we investigate in which terms liquidity crises may trigger large price fluctuations at smaller time scales. We will consider time windows and returns taking \textit{events} of $\Delta t=30$ seconds and the state of the LOB just before them. Practically, we study the condition of the LOB at the moment of the last operation before the beginning of the time window which defines the event. We point out that in this way we neglect the orders' flow inside the time windows, adopting a different approach with respect to the analysis we performed in the previous sections. In 
fact, we 
believe that on smaller time scales the static structure of the book is the key element to determine the magnitude and the sign of price movements.\\ The smallness of the considered time intervals permits to study the condition of the LOB before the large price fluctuations with a significant number of events at our disposal. First 
of all we check if some average property of the LOB is changed before the large events. A natural observable is the LOB profile, that is, the volume present in the form of available LOs as a function of the distance, expressed in price units or ticks, from the respective bests, usually averaged over a certain time interval\cite{Challet2001285}. Bouchaud et al. \cite{Bouchaud:2002} have found that the LOB profile has an universal shape, even if large temporal fluctuations are present. The study of these fluctuations concerns, using the terminology of Sarr and Lybek \cite{sarr2002measuring}, the breadth and the depth of the order book, instead of its resilience, which is a property closer to the ones we analyzed in the previous section. In order to illustrate how the fluctuations of the order book profile can influence the magnitude of price movements we consider a real situation, showed in Fig.\ref{sfondexample}, and in particular we focus on the structure of the LOB before (left) and after (right) the arrival of a MO. Before the 
transaction, we have a different breadth of the two sides of the book. In fact, while the bid side is dense and able to absorb a discrete amount of incoming MOs without affecting the midprice in a relevant way, the ask side is far more fragile, because the presence of large gaps between the order indicate that the arrival of sufficiently large MOs could trigger large movements of the midprice, that is what happens just a second later (figure on the right).\\
In order to explore this idea in a systematic way, we select the large events with the same methodology we used before, but considering $\Delta t=30$ seconds, and $x=0.3\%$ and $n=6\sigma$ as thresholds for the absolute and relative filters, respectively. Furthermore, we discard those events which are preceded by another large event within 90 seconds. Again, we consider 
only large positive events, but our results are qualitatively similar for the negative ones. In Fig.\ref{OBdepl} we plot the average profile (dashed lines) of the two sides of the LOB in comparison with the same sides just before the large positive fluctuations, that is, their state at the beginning of the considered time window (solid lines). One can see that, while the pressing side (which is, for positive returns, the bid side, red lines) remains substantially unchanged, the side which is about to break down (in this case the ask, blue lines) is characterized by substantially lower volumes, especially near the best. We can state that there is an intrinsic instability of that side of the book, that can cause large price jumps even if the incoming flow of MO is not significantly higher than the average. In the following section we introduce a quantitative framework to measure the breadth and the depth of the book and the correlation between the lack of liquidity and the ensuing returns.
\subsection{A microstructural definition of liquidity}
\begin{figure}[!ht]
\centering
\includegraphics[scale=0.35]{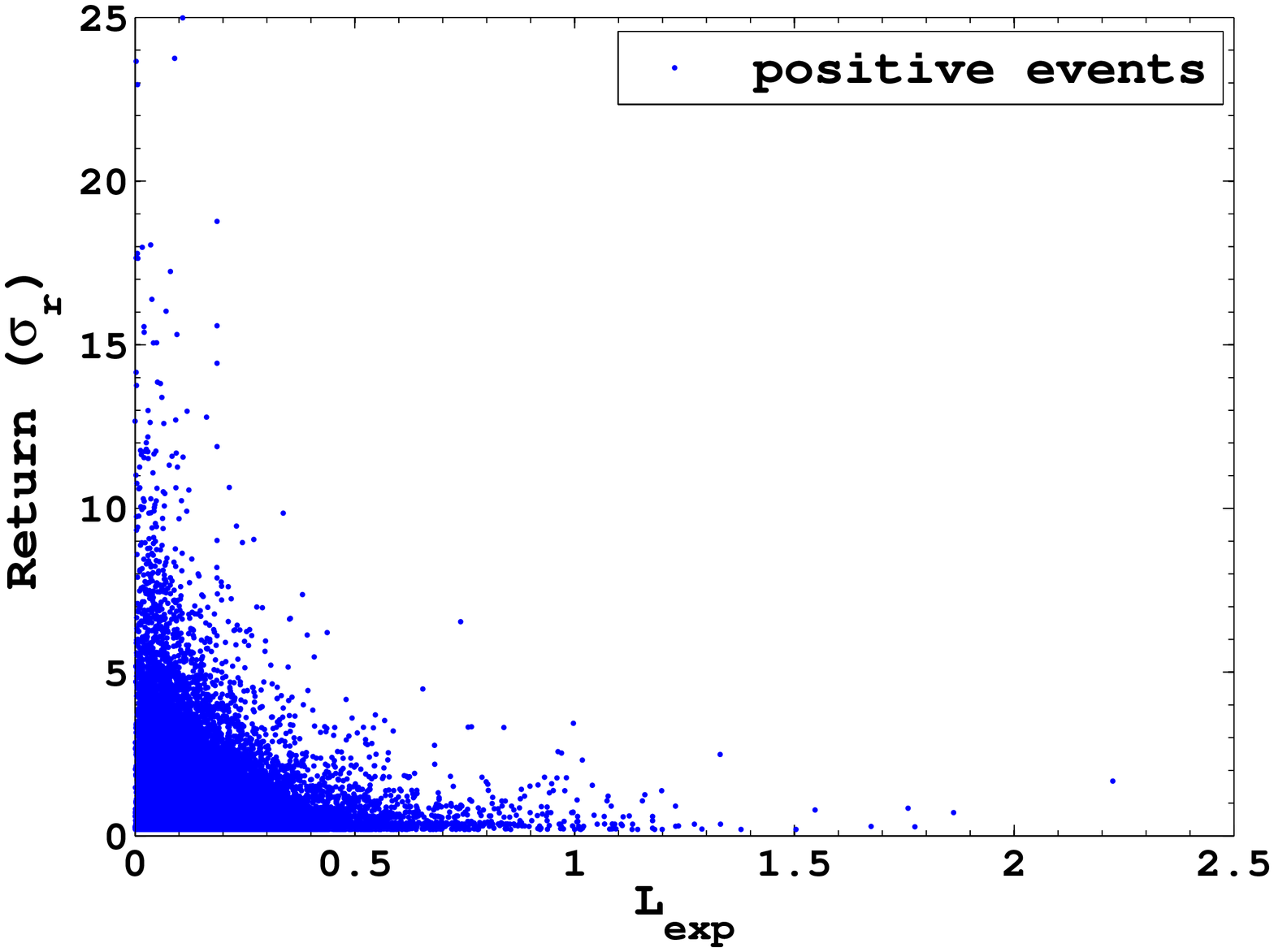}
\includegraphics[scale=0.3]{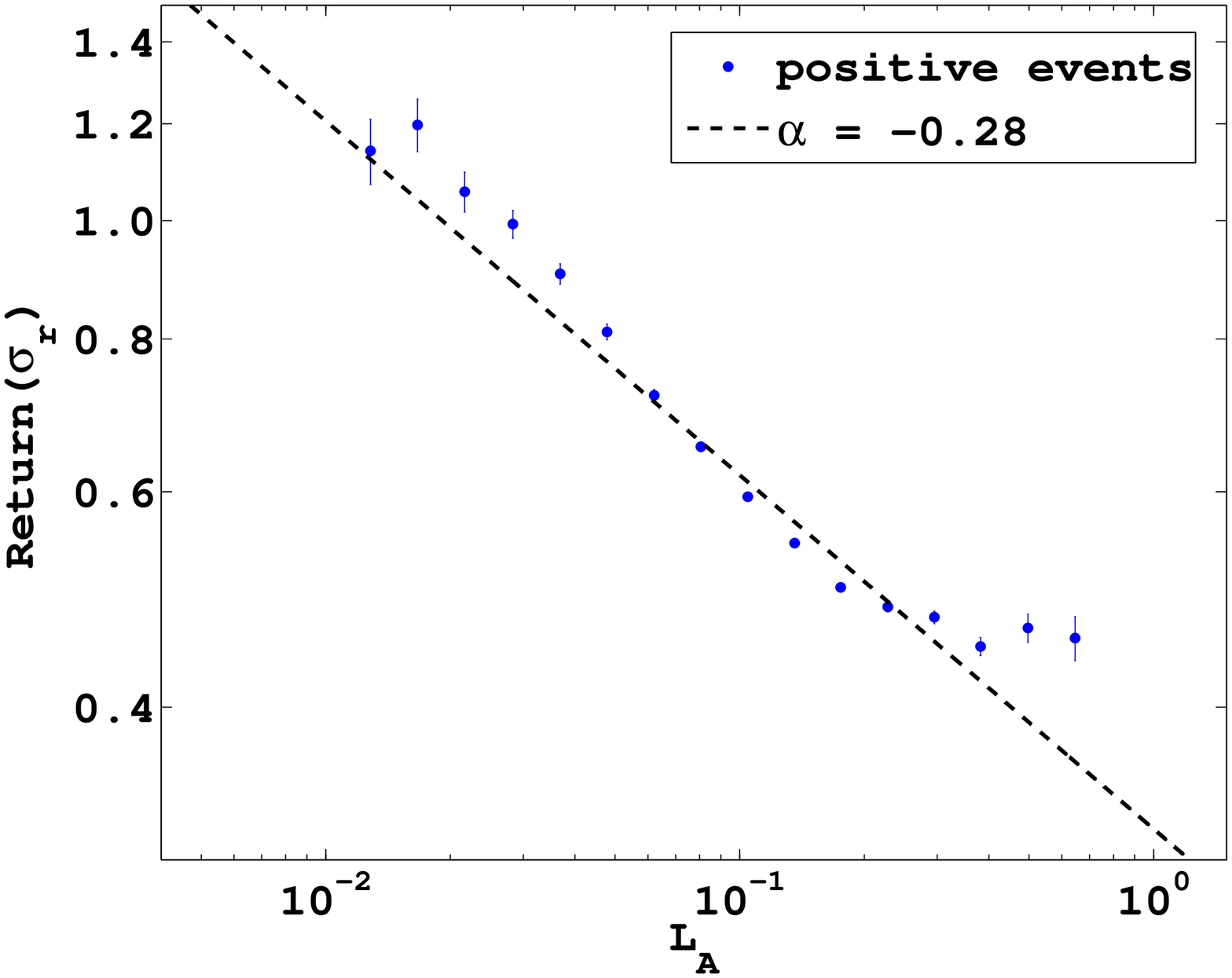}
\caption{Left: Returns, measured as multiples of the average standard deviation, and liquidity of the ask side of the book for all time windows with positive returns. High returns are present most probably during liquidity crises. Right: Power law fit and binned version of the cloud of points. The clear deviation at high values of liquidity is due to the presence of a minimum value for the possible returns.}
\label{nuvolaefit}
\end{figure}

\begin{figure}[!ht]
\centering
\includegraphics[scale=0.55]{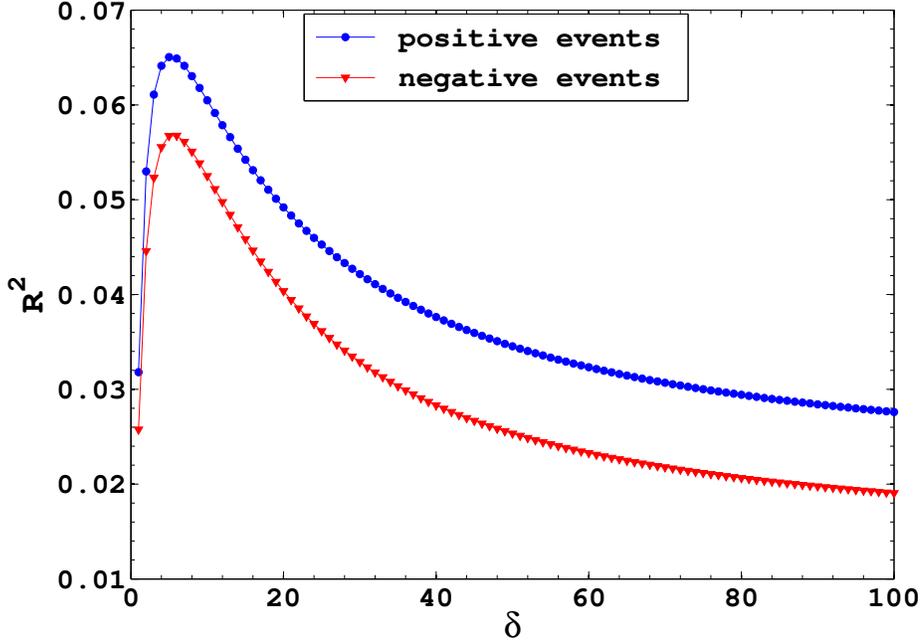}
\caption{Coefficient of determination relative to the power law fit of returns versus liquidity for positive (blue dots) and negative (red triangles) events, as a function of the parameter $\delta$, the characteristic distance of the exponential liquidity. The correlation is higher for values of $\delta$ around 5-6.}
\label{RsqVsDelta}
\end{figure}

As already noted in \cite{farmer6,farmer8}, the presence of gaps in the LOB plays a key role in enhancing the impact of MOs on the price. In particular, the authors have shown that the distribution of returns is to a large extent coincident with the one of the gaps, if one considers placement of single orders. However, when different time scales are considered, also the amount of shares and their distribution at different prices must be taken into account. In order to cope with these needs we study the breadth and depth of the LOB, which we quantify calculating the \textit{exponential liquidity} 
\begin{equation}
 L(\delta)=\frac{1}{<V_{N}>}\sum_{\Delta=1}^{N} V(\Delta) \exp(-\Delta / \delta)
\label{lexp}
 \end{equation}
where $<V_{N}>$ is the average volume in the book present within a maximum distance of $N=100$ ticks from the best, $V(\Delta)$ is the volume placed at tick $\Delta$, where $\Delta=1$ corresponds to the best price, and $\delta$ is a parameter which regulates the exponential weights, defining the relative importance of the orders. The liquidity is calculated on one side of the book and at the beginning of the time window. In the following we will indicate the liquidity which is computed on the ask side with $L_A$ and the one computed on the bid side with $L_B$. To study the correlation between the state of the LOB and the following return we plot the latter and the ask-side liquidity for all the time windows containing positive returns. The results are presented in Fig.\ref{nuvolaefit}(left), where each point corresponds to a different time window of 30 seconds and the returns are expressed in units of the average standard deviation of the year. The same analysis can be performed using negative events and the bid side of the book, obtaining analogous results. 
One can see that the number of events prevents a visual inspection of the core and of the distribution of the resulting cloud of points. However, it is evident that high returns occur only if liquidity is low and high values of liquidity prevent large fluctuations. In order to investigate this result in a quantitative way we fit the whole set of points with a power law. We compare this fit with a logarithmic binning in Fig.\ref{nuvolaefit}(right). The deviation for small return values is due to the presence of a minimum value of the possible returns, which is a consequence of the discrete nature of the order book. The minimum return is equal to one tick; this effect is less evident, but present, also in the figure on the right.\\
We have verified that these results are stable for different values of the parameters in Eq.(\ref{lexp}), as long as $N$ is large enough. The characteristic distance $\delta$ can be chosen, for instance, to maximize the goodness of fit, as measured by $R^2$ \footnote{In Fig.\ref{nuvolaefit} we have used those values of exponential liquidity which corresponds to a choice of $\delta$ which maximizes $R^2$.}. In Fig.\ref{RsqVsDelta} we plot the values of $R^2$ we obtain when we use different values of $\delta$ to calculate the exponential liquidity. One can notice a maximum around $\delta=\delta^\star \approx 5-6$ and then a sharp decline. This is a clear evidence that, for the considered time scale, the volumes near the best influence more the subsequent returns, but all the book must be considered. In a further analysis, which we do not show for reasons of space, we have found that the value of $\delta^\star$ increases with the length of the time window $\Delta t$. Obviously, if $\Delta t$ 
is too large, the correlation disappears, because at longer time scales price fluctuations are mostly driven by order flow effects, studied in the previous sections, rather than a static depletion of the book (to give an idea, in a time window of 15 minutes hundreds of orders are placed or canceled).\\ At a first look the values of the correlation coefficient could seem too little to be meaningful. However, one has to consider that the number of points is large enough to make the results of our fit statistically significant, as can be seen if one performs a t-test and checks the resulting p-value. A direct consequence is the low errors associated to the estimated parameters. Assuming a relationship of the kind $r(t)=K L_{A}^{-\alpha}$ one finds $K=0.2944 \pm 0.0028$ and $\alpha=−0.2800 \pm 0.0041$ for positive events. The analysis of the negative events gives values of $A$ and $\alpha$ compatible within error bars.\\ Recently Yura et al.\cite{PhysRevLett.112.098703} introduced a new mathematical framework 
to investigate the microstructural dynamics of stock prices, seen as colloids which interact with the various layers of the LOB. 
In this work we have found qualitatively compatible 
results, namely, that orders placed near the best should weigh more. In particular, they use larger time windows (on average) and find the largest correlation at a higher distance from the best with respect to our $\delta^\star \approx 5-6$. Even if a systematic comparison is made difficult by the different datasets we use and by the different definitions of both time windows and liquidity, since we observe an increase of $\delta^\star$ with $\Delta t$ we can reasonably expect the two analyses to be compatible.\\
We end this section noticing that, in general, many possible empirical measures of liquidity are possible. According Aitken and Comerton-Forde, the liquidity measures used in literature are actually 68 (see \cite{Aitken200345} and references therein), but many of them are widely discussed, leading Goyenko et al. to the choice of a meaningful title for their review of the subject: ``Do liquidity measures measure liquidity?'' \cite{Goyenko2009153}. A systematic comparison of some measures of liquidity with the ones we introduce in this article will be presented in a forthcoming work.
\subsection{Liquidity imbalance}
\begin{figure}[!ht]
\centering
\includegraphics[scale=0.5]{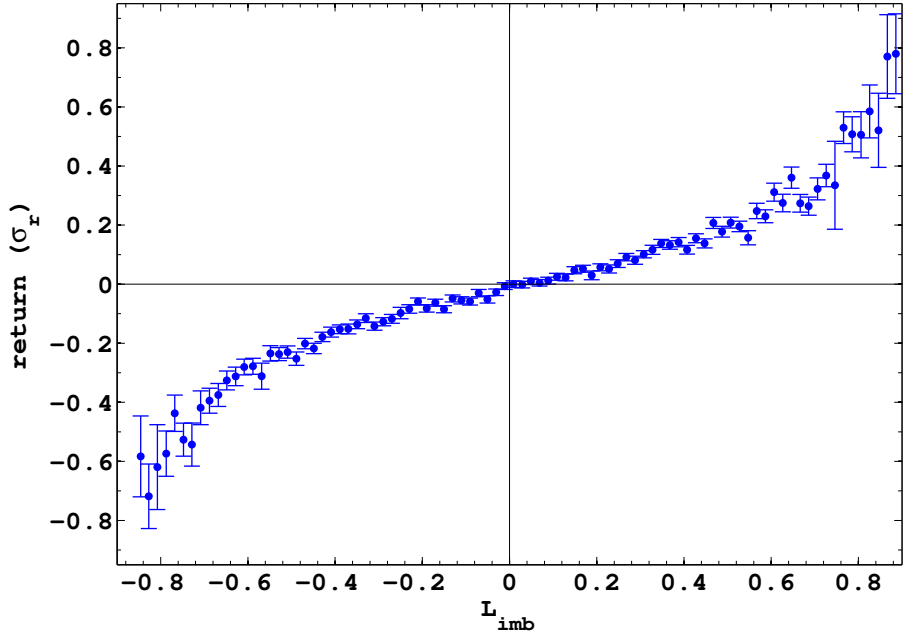}
\caption{Average return given the liquidity imbalance. The return is measured over all the time windows (which are 30 seconds long), and the imbalance at the beginning of them. We find a monotonic, non linear dependence.}
\label{RetLimb}
\end{figure}

\begin{figure}[!ht]
\centering
\includegraphics[scale=0.55]{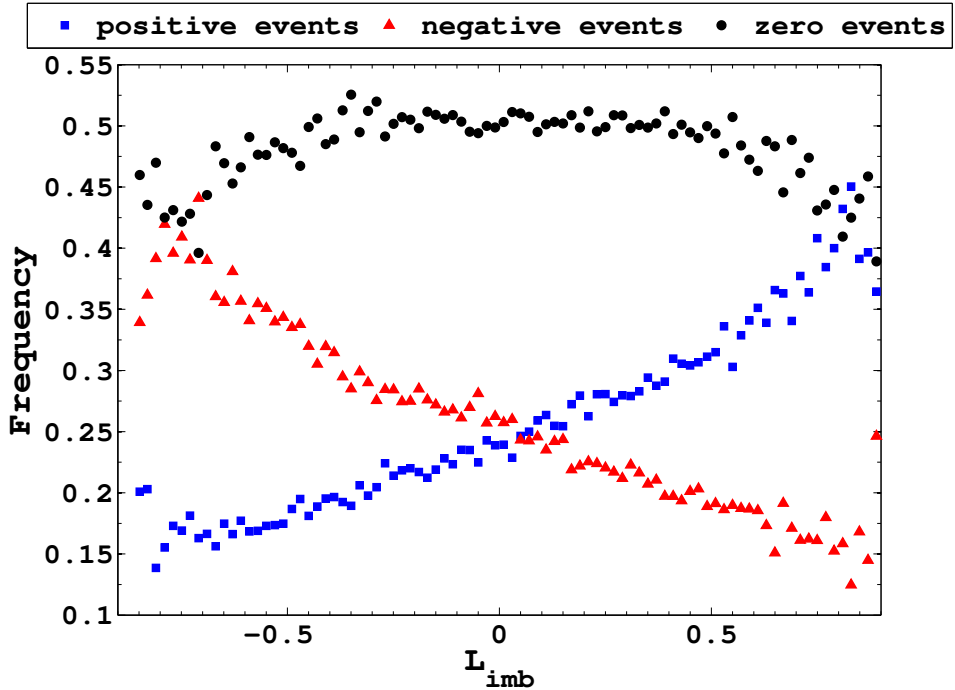}
\caption{Relative frequencies of return signs as a function of the liquidity imbalance defined by Eq.(\ref{limb}). The imbalance is calculated at the beginning of the time window which defines the price movement. Positive liquidity imbalances are often associated with positive returns, and viceversa. For instance, when $L_{imb} \approx 0.8$, one finds a number of positive returns which is roughly double than the number of the negative ones.}
\label{pmz}
\end{figure}

In the previous section we measured the liquidity considering only one side of the book and conditioning our analysis to the presence of a future return of a given sign. As a consequence, we left aside returns exactly equal to zero (which are, especially for small time windows, a considerable fraction of the total) and we did not study any possible predictive power on the sign which this kind of approach could have. In order to define a microstructural property of the book which takes into account the static asymmetry between the volumes placed on both sides we define the \textit{liquidity imbalance} as 
\begin{equation}
 L_{imb}(\delta)=\frac{L_{B}(\delta)-L_{A}(\delta)}{L_{B}(\delta)+L_{A}(\delta)}
\label{limb}
 \end{equation}
where $L_{A}(\delta)$ is the exponential liquidity defined in Eq.(\ref{lexp}) and calculated for the ask side, and $L_B(\delta)$ refers instead to the bid side. As before, we measure these quantities just before the beginning of the time window. Given the definition of Eq.(\ref{limb}), we expect that the state of the book facilitates positive returns when $L_{imb}(\delta)$ is positive, and viceversa. If one takes $\Delta t=30$ seconds and $\delta=5$ ticks one obtains a cloud of points in the $L_{imb}$-returns plane. A non parametric fit of this set of points is depicted in Fig.\ref{RetLimb}. Each point represents the average over all the returns which occur on all the time windows which follow a given liquidity imbalance. The bars are given by two times the standard error of the average value. We remind that each time window is 30 seconds long and the liquidity is computed at the beginning of the time interval. One finds a clear, non linear correlation between the liquidity imbalance and the return in the 
following time window.\\
In Fig.\ref{pmz} we repeat the same analysis but considering the frequency of three possible outcomes for the return which follows the LOB state as measured by the liquidity imbalance: positive (blue squares), zero (black dots) and negative returns (red triangles). Even if about half of the events have zero return, it is evident that different states of the book lead to different frequencies of subsequent return signs. In particular, high values of $L_{imb}$ are clearly associated to positive returns. For example, if one takes a state of the book in which $L_{imb} \approx 0.8$, the frequency of the positive returns is more than double the frequency of the negative ones. To assess the predictive power of this kind of analysis one should measure this effect on a training set of events and then check its ability to forecast future price movements on a test set. We aim to perform this kind of test in a future work.
\section{Conclusions and discussion}
In this work we present an empirical study of the static and dynamic properties of the limit order book and their relationship with price movements. In particular, we study two effects, both connected to the concept of liquidity, which are present on different time scales. We divide our time series in time windows of length $\Delta t=15$ minutes and 30 seconds. We find that on (relatively) large time scales large price fluctuations are connected to a low resilience of the market, namely, a lack of response, in terms of limit orders, to an incoming flow of market orders. Obviously, at this stage of analysis we are not able to claim a direct cause-effect relationship between order flow imbalance and price jumps, because we do not have a time delay between the two and we do not perform the opposite analysis, that is, we do not consider all the events with a clear order flow imbalance. On smaller time scales, instead, we adopt a static approach, in which the state of the book at the beginning of the time window 
is 
analyzed. 
A suitable measure of liquidity, and in particular of the breadth and the depth of the market, is presented. We find that the liquidity present on one side of the book is correlated with the magnitude of the immediately following return. Moreover, the liquidity imbalance between the two sides of the book is clearly related with the sign of this return.\\
Such an analysis can be inserted in the wider field which studies the lack of linearity and simple cause-effect mechanisms in financial markets. In fact, we find that liquidity crises can enhance in a strongly non linear way the predisposition towards instability of markets. This behavior defines a stylized fact which has to be taken into account in both theoretical modeling and practical applications. In order to cope with these effects we suggest to consider measures of liquidity like the ones we proposed in practical situations, such as the evaluation of the expected price impact of an order. This can be accomplished, for example, by considering a two-dimensional price impact function, like the one introduced in \cite{nostroorderbook}, which explicitly takes into account the impact dependence on both volume and liquidity.

% tables should appear as floats within the text
% The ruledtabular enviroment adds doubled rules to table and sets a
% reasonable default table settings.
% Use the table* environment to get a full-width table in two-column
% Add \usepackage{longtable} and the longtable (or longtable*}
% environment for nicely formatted long tables. Or use the the [H]
% placement option to break a long table (with less control than 
% \begin{table}%[H] add [H] placement to break table across pages
% \begin{ruledtabular}
% Lines of table here ending with \\
% \end{tabular}
% \end{ruledtabular}
% \end{table}

%\appendix
%\section{}

\begin{acknowledgments}
We thank Diego Fiorani for the preparation of the data and Marco Bartolozzi, Matthieu Cristelli, and Andrea Crisanti for useful discussions. This work was supported by the European project FET-Open GROWTHCOM (grant num. 611272) and the Italian PNR project CRISIS-
Lab. The funders had no role in study design, data collection and analysis, decision to publish, or preparation of the manuscript.
\end{acknowledgments}

\bibliography{biblio}

\end{document}